\documentclass{llncs}

\usepackage{amssymb}
\usepackage{amsmath}
\usepackage{epic}
\usepackage{eepic}
\usepackage{graphicx}
\DeclareSymbolFontAlphabet{\mathbb}{AMSb}
\usepackage{Vmargin}
\setmargins{1in}{1in}{6.5in}{9in}{0pt}{0pt}{12pt}{42pt}
\usepackage{times}
\usepackage{amsfonts}
\usepackage{eucal}
\newtheorem{theo}{Theorem}

\begin{document}

\title{Capacity constraints and the inevitability of mediators in adword auctions}

\author{ Sudhir Kumar Singh\inst{1,}\thanks{This work was done while the author was working for Ilial Inc.. The financial
support from Ilial Inc. is highly acknowledged.} \and  Vwani P. Roychowdhury\inst{1,2}
 \and Himawan Gunadhi\inst{2} \and Behnam A. Rezaei\inst{2}}
\institute{ Department of Electrical Engineering, University of California, Los Angeles, CA 90095.\\
\and Ilial Inc., 11943 Montana Ave. Suite 200, Los Angeles, CA 90049. \\
{\it \{sudhir,vwani,gunadhi,behnam\} @ ilial.com}}

\date{\today}
\maketitle

\begin{abstract}

One natural constraint in the sponsored search advertising framework arises from the fact
that there is a limit on the number of available slots, especially for the popular
keywords, and as a result, a significant pool of advertisers are left out.  We study the
emergence of diversification in the adword market triggered by such capacity constraints
in the sense that new market mechanisms, as well as, new for-profit agents are likely to
emerge to combat or to make profit from the opportunities created by shortages in ad-space
inventory. We propose a model where the additional capacity is provided by  for-profit
agents (or, mediators), who compete for slots in the original auction, draw traffic, and
run their own sub-auctions. The quality of the additional capacity provided by a mediator
is measured by its {\it fitness} factor. We  compute revenues and payoffs for all the
different parties at a {\it symmetric Nash equilibrium} (SNE) when the mediator-based
model is operated by a mechanism currently being used by Google and Yahoo!, and then
compare these numbers with those obtained at a corresponding SNE for the same mechanism,
but without any mediators involved in the auctions. Such calculations allow us to
determine the value of the additional capacity. Our results show that the revenue of the
auctioneer, as well as the social value (i.e. efficiency ), always increase when mediators
are involved; moreover even the payoffs of {\em all} the bidders will increase if the
mediator has a high enough fitness. Thus, our analysis indicates that there are
significant opportunities for diversification in the internet economy and we should expect
it to continue to develop richer structure, with room for different types of agents and
mechanisms to coexist.

\end{abstract}

\section{Introduction}
Sponsored search advertising is a significant growth market and is witnessing rapid growth and
evolution. The analysis of the underlying models has so far primarily focused on the
scenario, where advertisers/bidders interact directly with the auctioneers, i.e., the
Search Engines and publishers. However, the market is already witnessing the spontaneous
emergence of several categories of companies who are trying to mediate or facilitate the
auction process. For example, a number of different AdNetworks have started proliferating,
and so have companies who specialize in reselling ad inventories. Hence, there is a need
for analyzing the impact of such incentive driven and for-profit agents, especially as
they become more sophisticated in playing the game. In the present work, our
focus is on the emergence of market mechanisms and for-profit agents motivated by
capacity constraint inherent to the present models.

For instance, one natural constraint comes from the fact that there is a limit on the
number of slots available for putting ads, especially for the popular keywords, and a
significant pool of advertisers are left out due to this capacity constraint. We ask
whether there are sustainable market constructs and mechanisms, where new players interact
with the existing auction mechanisms to increase the overall capacity.  In particular,
lead-generation companies who bid for keywords, draw traffic from search pages and then
redirect such traffic to service/product providers, have spontaneously emerged. However,
the incentive and equilibria properties of paid-search auctions in the presence of such
profit-driven players have not been explored. We investigate key questions, including what
happens to the overall revenue of the auctioneers when such mediators participate, what is
the payoff of a mediator and how does it dependent on her quality, how are the payoffs
of the bidders affected, and is there an overall value that is generated by such
mechanisms.

Formally, in the current models, there are $K$ slots to be allocated among  $N$ ($\geq K$)
bidders (i.e. the advertisers). A bidder $i$ has a true valuation $v_i$ (known only to the
bidder $i$) for the specific keyword and she bids $b_i$. The expected {\it click through
rate} (CTR) of an ad put by bidder $i$ when allocated slot $j$ has the form $\gamma_j e_i$
i.e. separable in to a position effect and an advertiser effect. $\gamma_j$'s can be
interpreted as the probability that an ad will be noticed when put in slot $j$ and it is
assumed that $\gamma_1 > \gamma_2
>\dots > \gamma_K >  \gamma_{K+1}=  \gamma_{K+2} = \dots \gamma_{N} = 0$. $e_i$ can be interpreted as the probability that an ad put by
bidder $i$ will be clicked on if noticed and is refered as the {\it relevance} of bidder
$i$. The
payoff/utility of bidder $i$ when given slot $j$ at a price of $p$ per click is given by
$e_i\gamma_j (v_i - p)$ and they are assumed to be rational agents trying to maximize
their payoffs. As of now, Google as well as Yahoo! uses schemes closely modeled as RBR(rank by revenue)
with GSP(generalized second pricing). The bidders are ranked according to $e_iv_i$ and the slots are allocated as per
this ranks. For simplicity of notation, assume that the $i$th bidder is the one allocated
slot $i$ according to this ranking rule, then $i$ is charged an amount equal to
$\frac{e_{i+1} v_{i+1}}{e_i}$.
Formal analysis of such sponsored search advertising model has been done extensively in recent years, from
algorithmic as well as from game theoretic perspective\cite{EOS05,MSVV05,Lah06,AGM06,Var06,LP07,MNS07}.

In the following section, we propose and study a model  wherein the additional capacity is
provided by a for-profit agent who competes for a slot in the original auction, draws
traffic and runs its own sub-auction for the added slots. We discuss the cost or the value
of capacity by analyzing the change in the revenues due to added capacity as compared to
the ones without added capacity.

\section{The Model}

In this section, we discuss our model motivated by the capacity constraint,
which can be formally described as follows:

\begin{itemize}

%\item The additional capacity (i.e. the additional slots) is provided by third parties called {\it mediators}.
%and not the auctioneer/search engine.

\item {\bf Primary Auction ($p$-auction) :}  Mediators participate in the original auction run by the search engine (called {\it $p$-auction}) and
compete with advertisers for slots (called {\it  primary slots}).
For the $i$th agent (an advertiser or a mediator), let $v_i^p$ and $b_i^p$ denote
her true valuation and the bid for the $p$-auction respectively.
Further, let us denote $v_i^p e_i^p$ by $s_i^p$ where $e_i^p$ is the relevance score of $i$th agent for $p$-auction.
Let there are $\kappa$ mediators and there indices are $M_1, M_2, \dots, M_{\kappa}$ respectively.  \\

\item {\bf Secondary auctions ($s$-auctions):}

\begin{itemize}

\item {\bf Secondary slots:} Suppose that in the primary auction, the slots assigned to the mediators are $l_1,l_2,\dots,l_{\kappa}$ respectively,
then effectively, the additional slots are obtained by forking these {\it primary slots}
in to $L_1,L_2,\dots, L_{\kappa}$ additional slots respectively, where $L_i \leq K$ for all $i=1,2,\dots, \kappa$.
By forking we mean the following: on the associated landing page the mediator puts some information
relevant to the specific keyword associated with the $p$-auction
along with the space for additional slots.
Let us call these additional
slots as {\em secondary slots}. \\

\item {\bf Properties of secondary slots and {\it fitness} of the mediators:} For the $i$th mediator,
there will be a probability associated with her ad to be clicked if noticed, which is
actually her relevence score $e_{M_i}^p$ and the position based CTRs might actually
improve say by a factor of $\alpha_i$. This means that the position based CTR for the
$j$th secondary slot of $i$th mediator in modeled as $\alpha_i \gamma_j$ for $1 \leq j
\leq L_i$ and $0$ otherwise. Therefore, we can define a {\it fitness} $f_i$ for the $i$th
mediator, which is equal to $e_{M_i}^p \alpha_i$. Thus corresponding to the $l_i$th
primary slot (the one being forked by the $i$th mediator), the {\it effective} position
based CTR for the $j$th secondary slot obtained is $\tilde{\gamma}_{i,j}$ where
\begin{equation}
\tilde{\gamma}_{i,j} =  \left\{  \begin{array}{ll}\gamma_{l_i} f_i \gamma_j & \textrm{ for $j=1,2,\dots, L_i$,  } \\
   0  &  \textrm{  otherwise. }
\end{array} \right. \label{eq:effec-ctr}
\end{equation}
 Note that $ f_i \gamma_1 < 1$,
however $f_i$ could be greater than $1$. \\

\item {\bf $s$-auctions: } Mediators run their individual sub-auctions (called {\it $s$-auctions})
for the secondary slots provided by them.
For an advertiser there is another type of valuations and bids, the ones associated with $s$-auctions.
For the $i$th agent, let $v_{i,j}^s$ and $b_{i,j}^s$ denote
her true valuation and the bid for the $s$-auction of $j$th mediator respectively.
In general, the two types of valuations or bids corresponding to $p$-auction and the $s$-auctions might
differ a lot.
We also assume that $v_{i,j}^s=0$ and $b_{i,j}^s=0$ whenever $i$ is a mediator.
Further, for the advertisers who do not participate in one auction ($p$-auction or $s$-auction), the
corresponding true valuation and the bid are assumed to be zero.
Also, for notational convenience let us denote
 $v_{i,j}^s e_{i,j}^s$ by $s_{i,j}^s$ where $e_{i,j}^s$ is the relevance score of $i$th agent for the
$s$-auction of $j$th mediator.  \\

\item {\bf Payment models for $s$-auctions:}  Mediators could sell their secondary slots by impression (PPM), by pay-per-click (PPC) or pay-per-conversion(PPA).
In the following analysis, we consider PPC. \\

\end{itemize}

\item {\bf Freedom of participation:} Advertisers are free to bid for primary as well as secondary slots.   \\

\item {\bf True valuations of the mediators:}
The true valuation of the mediators are derived from the expected revenue (total payments from advertisers)
they obtain from the corresponding
$s$-auctions\footnote[1]{This way of deriving the true valuation for the mediator is reasonable for the mediator can participate
in the $p$-auction several times and run her corresponding $s$-auction and can estimate the revenue
she is deriving from the $s$-auction.} {\it ex ante}.

\end{itemize}

\section{Bid Profiles at SNE}

For simplicity, let us assume participation of a single mediator and the analysis involving several mediators
can be done in a similar fashion.
For notational convenience let
\begin{align*}
f=f_1,  \textrm{ the fitness of the mediator} \\
l =l_1, \textrm{ the position of the primary slot assigned to the mediator } \\
L=L_1 , \textrm{ the number of secondary slots provided by the mediator in her $s$-auction} \\
M=M_1 , \textrm{ the index of the mediator i.e. $M$th agent is the mediator } \\
\tilde{\gamma}_j = \tilde{\gamma}_{1,j}, \textrm{ is the {\it effective} position based CTR of the $j$th secondary slot provided by the mediator}  \\
v_{i,1}^s = v_i^s , \textrm{  is the true valuation of the agent $i$ for the $s$-auction} \\
b_{i,1}^s = b_i^s, \textrm{  is the bid of the agent $i$ for the $s$-auction, and } \\
s_{i,1}^s = s_i^s =  v_i^s e_i^s , \textrm{ where  $e_i^s=e_{i,1}^s$ is the relevance score of $i$th agent for the $s$-auction}.
\end{align*}

The $p$-auction as well as the $s$-auction is done via {\em RBR} with {\em GSP}, i.e. the mechanism
currently being used by Google and Yahoo!, and
the solution concept we use is {\em Symmetric Nash Equilibria(SNE)}\cite{EOS05,Var06}.
Suppose the allocations for the $p$-auction and $s$-auction are
$\sigma:\{1,2,\dots,N\} \longrightarrow \{1,2,\dots,N\} $ and $\tau:\{1,2,\dots,N\} \longrightarrow \{1,2,\dots,N\}$
respectively. Then the payoff of the $i$th agent from the combined auction ($p$-auction and $s$-auction together) is

\begin{displaymath}
u_i=\gamma_{\sigma^{-1}(i)} \left(s_i^p - r^p_{\sigma^{-1}(i)+1}\right) + \tilde{\gamma}_{\tau^{-1}(i)} \left(s_i^s - r^s_{\tau^{-1}(i)+1}\right)
\end{displaymath}
where
\begin{align*}
r_j^p=b_{\sigma(j)}^p e_{\sigma(j)}^p, \\
 r_j^s=b_{\tau(j)}^s e_{\tau(j)}^s.
\end{align*}
From the mathematical structure of payoffs and strategies available to the bidders wherein two different
uncorrelated values can be reported as bids in the two types of auctions independently of each other\footnote[2]{This assumption was motivated by some empirical examples
from Google Adword\footnotemark[3].},
it is clear that the equilibrium of the combined auction game is the one obtained from the
equilibria of the $p$-auction game and the $s$-auction game each played in isolation. In particular at {\em SNE}\cite{EOS05,Var06},
\begin{align*}
\gamma_i r_{i+1}^p = \sum_{j=i}^K (\gamma_j - \gamma_{j+1})  s_{\sigma(j+1)}^p \textrm {  for all $i =1,2,\dots, K$  }
\end{align*}
and
\begin{align*}
\tilde{\gamma}_i r_{i+1}^s = \sum_{j=i}^L (\tilde{\gamma}_j - \tilde{\gamma}_{j+1})  s_{\tau(j+1)}^s \textrm {  for all $i =1,2,\dots, L$  }
\end{align*}
which implies that (see Eq.~(\ref{eq:effec-ctr}))
\begin{align*}
\gamma_i r_{i+1}^s = \sum_{j=i}^{L-1} (\gamma_j - \gamma_{j+1})  s_{\tau(j+1)}^s + \gamma_L s_{\tau(L+1)}^s \textrm {  for all $i =1,2,\dots, L$  }
\end{align*}
where
\begin{align*}
s_{\sigma(l)}^p = s_M^p = f \sum_{j=1}^L \gamma_j r_{j+1}^s = f \left( \sum_{j=1}^{L-1} (\gamma_j - \gamma_{j+1}) j s_{\tau(j+1)}^s + \gamma_L L s_{\tau(L+1)}^s\right)
\end{align*}
is the true valuation of the mediator multiplied  by her relevance score as per our
definition\footnotemark[1], which is the expected revenue she derives from her $s$-auction
{\it ex ante} given a slot in the $p$-auction and therefore the mediator's payoff at SNE
is

\begin{displaymath}
u_M = \gamma_{l} f \left( \sum_{j=1}^{L-1} (\gamma_j - \gamma_{j+1}) j s_{\tau(j+1)}^s + \gamma_L L s_{\tau(L+1)}^s\right)  - \sum_{j=l}^K (\gamma_j - \gamma_{j+1})  s_{\sigma(j+1)}^p.
\end{displaymath}

\section{Revenue of the Auctioneer}

In this section, we discuss the change in the revenue of the auctioneer due to the involvement of the mediator.
The revenue of the auctioneer with the participation of the mediator is
\begin{displaymath}
R= \sum_{j=1}^K\gamma_j r_{j+1}^p = \sum_{j=1}^K (\gamma_j - \gamma_{j+1}) j s_{\sigma(j+1)}^p
\end{displaymath}
and similarly, the revenue of the auctioneer without the participation of the mediator is
\begin{displaymath}
\begin{array}{ll}
R_0 & = \sum_{j=1}^K (\gamma_j - \gamma_{j+1}) j s_{\tilde{\sigma}(j+1)}^p  \textrm{  where $\tilde{\sigma}(j) = \sigma(j)$ for $j < l$ and
$\tilde{\sigma}(j) = \sigma(j+1)$ for $j \geq l$ }  \\
& \\
& = \sum_{j=1}^{l-2} (\gamma_j - \gamma_{j+1}) j s_{\sigma(j+1)}^p + \sum_{j=l-1}^K (\gamma_j - \gamma_{j+1}) j s_{\sigma(j+2)}^p.
\end{array}
\end{displaymath}
Therefore,
\begin{align*}
R - R_0 & = \sum_{j=max \{1, l-1\}}^K (\gamma_j - \gamma_{j+1}) j (s_{\sigma(j+1)}^p-s_{\sigma(j+2)}^p) \\
& \geq 0  \textrm{ as } s_{\sigma(i)}^p \geq  s_{\sigma(i+1)}^p  \forall i=1,2,\dots, K+1  \textrm{ at {\em SNE}. }
\end{align*}

Thus revenue of the auctioneer always increases by the involvement of the mediator.
As we can note from the above expression, smaller the $l$ better the improvement in the revenue
of the auctioneer. To ensure a smaller value of $l$, the mediator's valuation which is the expected payments that she obtains
from  the $s$-auction should be better, therefore fitness factor $f$ should be very good. There is another way to improve her
true valuation. The mediator could actually run many subauctions related to the specific keyword in question.
This can be done as follows: besides providing the additional slots on the landing page, the information
section of the page could contain links to other pages wherein further additional slots associated with
a related keyword could
be provided\footnote[3]{For example, the keyword ``personal loans'' or ``easy loans'' and the mediator ``personalloans.com''.}.
With this variation of the model, a better value of $l$ could possibly be ensured
leading to a win-win situation for everyone.

\begin{theo}
Increasing the capacity via mediator improves the revenue of auctioneer.
\end{theo}

\section{Efficiency}

Now let us turn our attention to the change in the efficiency and as we will prove below, the efficiency always improves
by the participation of the mediator.

\begin{align*}
E_0 = \sum_{j=1}^K \gamma_j s_{\tilde{\sigma}(j)}^p = \sum_{j=1}^{l-1} \gamma_j s_{\sigma(j)}^p + \sum_{j=l}^K \gamma_j s_{\sigma(j+1)}^p \textrm{ and } \\
E =  \sum_{j=1}^{l-1} \gamma_j s_{\sigma(j)}^p + \sum_{j=l+1}^K \gamma_j s_{\sigma(j)}^p + \gamma_{l} f \sum_{j=1}^L \gamma_j s_{\tau(j)}^s
\end{align*}

\begin{align*}
\therefore E-E_0 & =  \gamma_{l} f \sum_{j=1}^L \gamma_j s_{\tau(j)}^s - \sum_{l}^K (\gamma_j - \gamma_{j+1}) s_{\sigma(j+1)}^p \\
& = \gamma_{l} f \sum_{j=1}^L \gamma_j s_{\tau(j)}^s - \gamma_{l} r_{l+1}^p \\
& \geq 0 \\
& \textrm{ as }  \gamma_{l} f \sum_{j=1}^L \gamma_j s_{\tau(j)}^s \geq \gamma_{l} f \sum_{j=1}^L \gamma_j r_{j+1}^s = \gamma_{l} s_{\sigma(l)}^p \geq \gamma_{l} r_{l+1}^p  \textrm{ at {\em SNE }.} \\
\end{align*}

\begin{theo}
Increasing the capacity via mediator improves the efficiency.
\end{theo}

\section{Advertisers' Payoffs}
Clearly,
for the newly accommodated advertisers, that is the ones who lost in the $p$-auction but
win a slot in $s$-auction, the payoffs increase from zero to a postitive number.
Now let us see where do these improvements in the revenue of the auctioneer, in payoffs of newly
accommodated advertisers, and in the efficiency
come from?
Only thing left to look at is the change in the payoffs for the advertisers who originally won in the $p$-auction,
that is the winners when there was no mediator.
The new payoff for $j$th ranked advertiser in $p$-auction is
\begin{displaymath}
u_{\sigma(j)} = \gamma_j s_{\sigma(j)}^p - \sum_{i=j}^K (\gamma_i - \gamma_{i+1})  s_{\sigma(i+1)}^p + u_{\sigma(j)}^s
\end{displaymath}

where
\begin{displaymath}
 u_{\sigma(j)}^s = \gamma_{l}f  \gamma_{\tau^{-1}(\sigma(j))} \left( s_{\sigma(j)}^s - r^s_{\tau^{-1}(\sigma(j))+1} \right)
\end{displaymath}
is her payoff from the $s$-auction.
Also, for $j \leq l-1$,  her payoff when there was no mediator is

\begin{displaymath}
\begin{array}{l}
u_{\sigma(j)}^0 = \gamma_j s_{\sigma(j)}^p - \sum_{i=j}^K (\gamma_i - \gamma_{i+1})  s_{\tilde{\sigma}(i+1)}^p \\
              \\
=  \gamma_j s_{\sigma(j)}^p - \sum_{i=j}^{l-2} (\gamma_i - \gamma_{i+1})  s_{\sigma(i+1)}^p -
\sum_{i=l-1}^K (\gamma_i - \gamma_{i+1})  s_{\sigma(i+2)}^p. \\
\\
\therefore u_{\sigma(j)} - u_{\sigma(j)}^0 =  u_{\sigma(j)}^s - \sum_{i=l-1}^K (\gamma_i - \gamma_{i+1}) (s_{\sigma(i+1)}^p- s_{\sigma(i+2)}^p)
\end{array}
\end{displaymath}

Similarly, for $ j \geq l+1$, her payoff when there was no mediator is
\begin{displaymath}
\begin{array}{l}
u_{\sigma(j)}^0 = \gamma_{j-1} s_{\sigma(j)}^p - \sum_{i=j-1}^K (\gamma_i - \gamma_{i+1})  s_{\sigma(i+2)}^p \\
\\
\therefore u_{\sigma(j)} - u_{\sigma(j)}^0 =  u_{\sigma(j)}^s - \sum_{i=j-1}^K (\gamma_i - \gamma_{i+1}) (s_{\sigma(i+1)}^p- s_{\sigma(i+2)}^p)
\end{array}
\end{displaymath}
Therefore, in general we have,
\begin{displaymath}
u_{\sigma(j)} - u_{\sigma(j)}^0 =  u_{\sigma(j)}^s - \sum_{i=max \{l-1,j-1\}}^K (\gamma_i - \gamma_{i+1}) (s_{\sigma(i+1)}^p- s_{\sigma(i+2)}^p).
\end{displaymath}

Thus, for the $j$th ranked winning advertiser from the auction without mediation, the revenue from the $p$-auction decreases by
$\sum_{i=max \{l-1,j-1\}}^K (\gamma_i - \gamma_{i+1}) (s_{\sigma(i+1)}^p- s_{\sigma(i+2)}^p)$ and she faces a loss unless
compensated for by her payoffs in $s$-auction.
Further, this payoff loss will be visible only to the advertisers
who joined the auction game before the mediator and they are likely to participate in the $s$-auction so as
to make up for this loss. Thus, via the mediator, a part of the payoffs of the originally winning advertisers
essentially gets distributed among the newly accommodated advertisers. However, when the mediator's fitness
factor $f$  is very good, it might be a win-win situation for everyone.
Depending on how good the fitness factor $f$ is, sometimes the payoff from the $s$-auction might be enough
to compensate for any loss by accommodating new advertisers.
Let us consider an extreme situation when $L=K$ and $\tau = \tilde{\sigma}$.
The {\em gain} in payoff for the advertiser $\sigma(j)$ is
\begin{displaymath}
\gamma_{l} f \sum_{i=j}^K (\gamma_i - \gamma_{i+1}) (s_{\sigma(j)}^s -s_{\sigma(i+1)}^s)  - \sum_{i=max \{l-1,j-1\}}^K (\gamma_i - \gamma_{i+1}) (s_{\sigma(i+1)}^p- s_{\sigma(i+2)}^p)
\end{displaymath}
Therefore as long as
\begin{displaymath}
f \geq  \frac{ \sum_{i=max \{l-1,j-1\}}^K (\gamma_i - \gamma_{i+1}) (s_{\sigma(i+1)}^p- s_{\sigma(i+2)}^p)}{\gamma_{l} \sum_{i=j}^K (\gamma_i - \gamma_{i+1}) (s_{\sigma(j)}^s -s_{\sigma(i+1)}^s) }
\end{displaymath}
the advertiser $\sigma(j)$ faces no net loss in payoff and might actually gain.

\section{Concluding Remarks}
In the present work, we have studied
the emergence of diversification in the adword market triggered by the inherent
capacity constraint.
We proposed and analyzed a model where additional capacity is created by a for-profit agent who compete
for a slot in the original auction, draws traffic and runs its own sub-auction.
Our study potentially indicate a $3$-fold diversification in the adword market in terms of
(i) the emergence of new market mechanisms,
(ii) emergence of new for-profit agents, and
(iii) involvement of a wider pool of advertisers.
Therefore,  we should expect the internet economy
to continue to develop richer structure, with room for different types of
agents and mechanisms to coexist.
In particular, capacity constraints motivates the study of
yet another model where the additional capacity is created by the search engine itself,
essentially acting as a mediator itself and running a single combined auction.
This study will be presented in an extended version of the present work.

\end{document}